\documentstyle[psfig]{l-aa}
\begin{document}
   \thesaurus{12         
              (11.03.4 A370 A2218;  
               11.05.2;  
               12.03.3;  
               12.07.1)} 

   \title{Number counts and redshift distribution of gravitational
arclets as a probe of galaxy evolution} 

   \author{J. B\'ezecourt, R. Pell\'o, G. Soucail}

   \offprints{J. B\'ezecourt, bezecour@obs-mip.fr}

   \institute{Observatoire Midi-Pyr\'en\'ees, Laboratoire d'Astrophysique, 
    UMR 5572, 14 Avenue E. Belin, F-31400 Toulouse, France }

\date{Received, Accepted }

\maketitle

\markboth{J. B\'ezecourt et al.: Number counts and redshift distribution
of gravitational arclets}{}

\begin{abstract}

We present a detailed model of the absolute number counts,
color and redshift distributions of gravitational arclets observed in
clusters of galaxies. The framework adopted for galaxy evolution is chosen
to fairly reproduce the observed number counts and redshift
distribution of field galaxies. Then, the spectrophotometric evolutionary
code is coupled with an accurate modelling of the cluster-lens mass
distribution. The interest in applying these
calculations to arclets is to use cluster-lenses as
filters to select faint distant galaxies. This
procedure is applied on two different cluster-lenses, Abell 2218 and
Abell 370, for which the mass distribution is well constrained.
We have studied the impact of the different sources of
uncertainty on the predicted number counts and redshift
distributions, taking into account the observational conditions for 
two sets of data, HST and ground-based images.
We investigate in details the influence of the mass modelling
on the counts and we show that simple cluster-scale potentials can 
no longer be used for arcs statistics.
The main result is that arcs at redshifts between 0.5 and 1 are
correctly predicted by the modelling as observed. Nevertheless,
an important population of high redshift arclets ($z \ge 1.0 $) is also
revealed by the simulations, which is not observed in spectroscopic
surveys of arclets. We discuss the nature of this disagreement,
probably due to uncertainties in the evolutionary models adopted here
for galaxies at high redshift. The spatial distribution of arclets 
in number density and the local mean redshift of the sample are also 
derived. These maps can be used as a tool to
optimize the search for high redshift galaxies magnified by the clusters
of galaxies.

\end{abstract}

\section{Introduction}
The measure of galaxy number counts and the study of the
spectrophotometric properties of faint galaxies are probably the
two leading topics to constrain the evolutionary history of galaxies,
a point which is absolutely needed in cosmology. The 
new observational tools coming from HST imagery surveys, deep 
number magnitude counts in different wavelengths and spectroscopic
surveys of faint galaxies have allowed to approach the 
spectromorphological evolution of galaxies in a reliable way. For
the first time, some clues have been proposed to explain the
nature of the faint blue galaxies and to determine the star
formation history (see the review by \cite{koo92} and 
\cite{ellis97}). Nevertheless, it is still 
difficult to estimate the rate at which this evolution occurs as well as the 
physical processes involved, because number counts are integrated values 
over the redshift distribution and the luminosity distribution of 
galaxies. Spectroscopic surveys in deep fields are one of the issues to 
treat the problem, but they are limited in magnitude. For example, the 
surveys up to the limit $B=24$ (\cite{glazebrook95}, \cite{cowie96})
show a peak at $z\simeq 0.5-0.7$, but the lack of a large number 
of high redshift objects in these surveys argues for a mild luminosity
evolution at intermediate redshift. From a deep sample of $I$-band selected
galaxies, the Canada France Redshift Survey (CFRS, \cite{lilly95}) 
reveals an evolution in the luminosity function (LF)
of the bluer field galaxies of about 1 magnitude between $z=0.7$ and $z=0$
but again fails to identify a population of high-$z$ galaxies. 
Spectroscopic data acquired by the Autofib Redshift Survey (\cite{ellis96})
leads to similar conclusions about the evolution of the luminosity 
function with redshift and show also 
a clear steepening in the faint end slope of the LF.

A powerful way to investigate higher redshifts with the same
magnitude limit is possible thanks to the gravitational magnification 
of background sources in the field of massive clusters of galaxies
(see \cite{fort94} for a review). 
Serendipitously lensed by a foreground cluster, gravitational arcs and 
arclets are representative of a rather distant population of galaxies, 
at least up to a redshift $\sim 1$ (\cite{soucail88}, \cite{pello91}, 
\cite{bezecourt97}, \cite{ebbels97}), or even higher in 
a few cases (\cite{mellier91}, \cite{ebbels96}). But again the 
magnitude limitation for spectroscopy prevents an extensive and deep 
analysis.
Photometric identification of very high redshift objects has also proved 
very successful in the case of multiple images lensed by a massive cluster 
of galaxies (\cite{trager97}, \cite{frye97}, \cite{pello97}).
To overcome the difficulty, several approaches are proposed, 
each of them being based on a redshift estimate for individual arclets.
A promising one is the lens inversion technique which consists 
in finding the most probable redshift for each arclet depending on its location
and shape parameters for a given lens model of the cluster potential 
(\cite{kneib94a}, \cite{kneib96}, \cite{ebbels97}). 
Successfully used in a few cases, it allows to reach 
intrinsic magnitudes up to $B \simeq$ 27. 
The redshift distribution of gravitational arclets can also be
estimated with multicolor photometry (\cite{pello96}), thanks to the 
photometric redshift estimated from a long wavelength baseline (from U to K). 
A more straightforward method
consists in measuring the depletion in counts of arclets caused by 
the amplification bias: depending on the slope of the field galaxies number
counts, the radial density of arclets will show a 
depletion curve characteristic of the sources redshift (\cite{broadhurst95}, 
\cite{fort96}). The drawing of the critical lines corresponding to the 
most distant galaxies was also proposed as a sub-product of this method 
and could be used to constrain the redshift of formation of the galaxies.

The aim of the paper is to probe the spectrophotometric evolution of
galaxies by computing both the number counts of arclets behind a cluster-lens 
and their redshift distribution. 
The basic idea is to consider cluster-lenses as ``high-$z$ filters'' which 
select background galaxies and distort them, making their shape 
easier to identify. So the sample of arclets is a subsample of 
a global sample of faint field galaxies in which galaxy evolution at 
relatively large look-back times may be much stronger. The deep number 
counts of arclets are then supposed to probe more directly the redshift 
distribution of {\em distant} galaxies and their relative weight with 
respect to a possible population of faint nearby galaxies. 
Previous statistical studies about the occurence rate 
of arcs or arclets assumed spherical potential for all 
clusters and/or neglected galaxy evolution (\cite{nemiroff89}, \cite{smail91}, 
\cite{wu93}, \cite{grossman94}, \cite{smail94},
\cite{refregier97}, \cite{hattori97}). The authors were more concerned about 
constraining the mass profile of the lenses, whereas in this paper we 
want to emphasize the implications on galaxy evolution. For this reason, 
we apply our models on real cluster-lenses (Abell 370 and Abell 2218)
for which the presence of multiple-images strongly constrains the
potential in the cluster core. Moreover in order to optimise the 
arclets identification among the crowded cluster fields, we use 
deep HST images available for both clusters. 

This paper is organised as follows: in Section 2, we compute the number counts 
of gravitational arclets by combining models of spectrophotometric evolution 
of galaxies, which reproduce the field number-counts, and realistic
models of mass distribution for clusters, which are well constrained by multiple 
images and giant arcs. The sensitivity of our model with respect to
uncertainties on the lensing potential at large radii is tested, and
the robustness of the results with respect to some observational criteria 
such as axis ratio or surface brightness is explored.
Section 3 presents the results obtained with this method on 
two cluster-lenses, A2218 and A370, including a comparison between the predicted 
number counts of arclets and the observed ones. The sensitivity of our model 
to different parameters is discussed in Section 4,
and some clues about the possibility to detect high-z galaxies are also
considered. Finally, conclusions are presented in Section 5.

Throughout the paper, we consider a Hubble constant of H$_0 = 50 
\, km \,s^{-1}\,Mpc^{-1}$, with $\Lambda = 0$. 

\section{Number counts and redshift distribution of gravitational
arclets}
The surface density of gravitational arclets obeys to two competing
effects: one is the magnification of the luminosity by the cluster 
potential and the other one is surface dilution.
If $n(<m)=n_0 10^{\alpha m}$ is the surface density of galaxies up to
magnitude $m$ with slope $\alpha$, the density of arclets magnified by a 
factor $M$ is
\begin{equation}
$$n_{arc}(<m)= {1\over M} \, n(< m + 2.5 log M)$$ 
\end{equation}
Hence,
\begin{equation}
$${n_{arc}(<m)\over n(<m)}= M^{2.5 \alpha -1}$$
\end{equation}
The dominating effect depends on
the slope of number counts in empty fields without magnification 
by an intervening mass. For steep counts 
($\alpha > 0.4$), gravitational lensing will increase the
surface density whereas for shallower counts a depletion will take
place (\cite{broadhurst95}, \cite{fort96}). 

Looking in details at the number counts behind a cluster-lens, we can write 
the number of arclets brighter than magnitude $m$ with an axis ratio
greater than $A_{min}$ and a surface brigthness brighter than $\mu_0$ 
within a given region of the sky as:
\begin{equation}
\begin{tabular}{ccc} 
$\lefteqn{N(m,A_{min},\mu_0) = }$ \\
 & & \\
 & & $ \sum_{i} \int_{z_l}^{z_{max}} \int_{A_{min}}^{\infty}
S(A,z) dA \ \int_{L_{min}}^{L_{max}} \Phi_i(L,z) \, dL \,  
{dV \over dz} \, dz $ \
\end{tabular}
\end{equation}
The summation is over the different morphological types and
$z_l$ is the lens redshift. $z_{max}(\mu_0,i)$ is the redshift cutoff
corresponding to the limit in central surface brightness $\mu_0$.
$S(A,z, H_0, \Omega)$ is the angular area in the source plane 
(at redshift $z$) filled by 
sources corresponding to arclets with an axis ratio between $A$ and $A+dA$.
$\Phi_i(L,z)$ is the LF at redshift $z$ for each
morphological type. 
$L_{min}(z,m,A, H_0, \Omega)$ is the luminosity of an object at redshift
$z$ which has an apparent magnitude $m$ after magnification by the cluster, 
assuming that the source is circular and that its axis ratio is A after
magnification (see below in \S 2.3). Finally, $L_{max}$ is the bright 
end of the LF and $dV(z,H_0, \Omega)$ is the volume element.

The main differences with standard number counts are:
firstly, the integration in redshift runs from the 
lens-redshift $z_l$ as a minimum up to a limit which depends essentially on 
the limitation in surface brightness  $z_{max}(\mu_0,i)$. Gravitational lensing
preserves the surface brightness of the sources, so a cut in observed surface 
brigthness is more realistic than a cut in magnitude for extended objects.
Secondly, one has 
to take into account the differential observed surface, computed in the source 
plane for each redshift and for each magnification. Finally, each 
luminosity has to be corrected from the magnification before the integration 
over the LF of the sources, as well as the minimum 
limit in luminosity. An additional application is to compute local values
of the different parameters 
to derive the 2D projected number density and the mean redshift of arclets
(see \S 3.6).

In order to validate this computation of number counts, we check first 
in the following section that observations in empty fields are correctly 
reproduced by the evolution model. We then present how to include a mass 
model in the problem, and the way both models are combined. We finally
investigate the sensivity of counts to these mass distributions.

\subsection{Number counts and redshift distribution of field galaxies}
\begin{table*}
\caption[]{Ingredients of the two number counts models, for two selected 
cosmologies.}
\label{tab-model}
\begin{flushleft}
\begin{tabular}{ccccccccccc}
\hline\noalign{\smallskip}
model&$q_0$&Spectral type&fraction&SFR&IMF&$z_f$&density&
$\Phi^{\ast}_i$&$\alpha$&$M_{B \ast}$\\
 & & & \% & & & &evolution&$10^{-3} h_{50}^3 Mpc^{-3}$& & \\
\noalign{\smallskip}
\hline\noalign{\smallskip}
1&0& & & & &4.5&no& & \\
 & &E/S0&0.28&exp. $\tau=1Gyr$& Scalo & & &0.95&--0.48&--20.87\\
 & &Sab/Sbc&0.47&exp. $\tau=10Gyr$& Scalo & & &1.15&--1.24&--21.14\\
 & &Scd/Sdm&0.22&constant& Salpeter & & &0.54&--1.24&--21.14\\
 & &vB&0.03& & Salpeter & & &0.12&--1.24&--21.14\\
2&0.5& & & & &5&yes& & & \\
 & &E/S0&0.28&exp. $\tau=1Gyr$& Scalo & & &0.95&--0.48&--20.87\\ 
 & &Sab/Sbc&0.47&exp. $\tau=8Gyr$& Scalo & & &1.15 &--1.24&--21.14\\
 & &Scd/Sdm&0.22&constant& Salpeter & & &0.54&--1.24&--21.14\\
 & &vB&0.03& & Salpeter & & &0.12&--1.24&--21.14\\
\noalign{\smallskip}
\hline
\end{tabular}
\end{flushleft}
\end{table*}

\subsubsection{Spectral energy distributions for template galaxies}
As a first step, one has to define a framework for galaxy evolution 
that fairly reproduce both the number counts and the redshift distribution 
of field galaxies. We follow here the results of Pozzetti et al. 
(1996, hereafter PBZ) using
the Bruzual and Charlot evolutionary code (1993 updated as GISSEL95).
Four galaxy types are used to represent the distribution in
morphological types. They correspond respectively to
an exponential star formation rate (SFR) for elliptical 
and spiral galaxies, with time scales of 1 Gyr and 10 Gyr for
$q_0=0$ (model 1) and 1 Gyr and 8 Gyr for $q_0=0.5$ (model 2). 
A constant SFR is assumed for late type galaxies and a 
population of eternally young objects is also introduced by the authors
to account for very blue objects. The redshift of formation for all galaxies 
is $z_{form} = 4.5$ for model 1 and 5 for model 2.
It should be noted that an additional assumption is made in PBZ
by considering different initial mass functions for
late type spirals (Salpeter IMF, \cite{salpeter55}) with respect to 
ellipticals and normal spirals
(Scalo IMF, Scalo 1986). This is necessary to obtain good fits to number counts
from $U$-band to $K$-band, although the
predicted colors for nearby galaxies does not match very well those observed 
and appear too red. 
A better agreement would require the use of a Salpeter IMF for all types but the
drawback is that too many high redshift galaxies are
produced, compared to spectroscopic surveys. Being aware of this discrepancy 
we choose as prime constraints number counts and redshift distribution of
field galaxies and adopt the same parameters as PBZ. A summary of 
the ingredients of our models is presented in Table \ref{tab-model}. The 
colors for the different galaxy types at $z=0$ are given in Table 
\ref{tab-colors}, and compared to those observed in the local universe
or directly derived from observed spectra (\cite{fukugita95}).

\begin{table*}
\caption[]{Colors of model and observed galaxies (\cite{fukugita95}).}
\label{tab-colors}
\begin{flushleft}
\begin{tabular}{cccccccccc}
\hline\noalign{\smallskip}
type & & $U-B$ & & & $B-V$ & & & $V-I$ & \\
 & $q_0=0$ & $q_0=0.5$ & observed & $q_0=0$ & $q_0=0.5$ & observed & 
$q_0=0$ & $q_0=0.5$ & observed \\
\noalign{\smallskip}
\hline\noalign{\smallskip}
E/S0 & 0.70 & 0.63 & 0.45/0.40 & 0.99 & 0.96 & 0.97/0.93 & 1.61 & 1.56 & 1.45/1.25 \\
Sab/Sbc & 0.22 & 0.19 & 0.13 & 0.68 & 0.65 & 0.73 & 1.33 & 1.29 & 1.30 \\
Scd/Sdm & --0.12 & --0.14 & --0.11 & 0.43 & 0.41 & 0.43 & 1.13 & 1.10 & 1.17 \\
vB & --0.61 & --0.61 & --- & 0.01 & 0.01 & --- & 0.62 & 0.62 & --- \\
\noalign{\smallskip}
\hline
\end{tabular}
\end{flushleft}
\end{table*}

\subsubsection{Luminosity function}
The weight affected to each galaxy 
type is taken from the determination of
the local LF by Efstathiou et al. (1988), with the
assumption that (1) morphological types are equivalent to spectroscopic types
at any redshift, and (2) the weights of spectromorphological types remain unchanged 
with $z$. A model fulfilling these two conditions is
a pure luminosity evolution model. Moreover, density evolution can occur due to
the addition of a population of dwarf objects at low redshift or to an earlier  
phase of merging. A change in the overall density is not required 
in a low $\Omega$ universe but it is necessary
in a closed universe to reproduce number counts (\cite{rocca90}, 
\cite{broadhurst92}). We consider here two different models, which correspond 
to these two cases:

\begin{itemize}

\item  Model 1: $q_0$=0, with a 
constant Schechter function, which is a pure luminosity evolution model.

\item Model 2:  $q_0$=0.5, a number luminosity evolution model which
keeps constant the comoving mass density. The LF evolves with z according to the 
merging law of Rocca--Volmerange and Guiderdoni (1990):
\begin{equation}
$$\Phi_i (L,z) =(1+z)^{2\eta} \, \Phi_i \left( L(1+z)^\eta ,z=0 \right)$$ 
\end{equation}
where 
\begin{equation}
$$\Phi_i (L,z=0) dL =\Phi^{\ast}_i \left( {L\over L_{\ast}(i)}
\right)^{\alpha(i)} e^{-{L\over L_{\ast}(i)}} {dL\over L_{\ast}(i)}$$ 
\end{equation}
is the LF at $z=0$ described by a Schechter law. 
$i$ corresponds to each morphological type and $\eta=1.5$ is adjusted 
by Rocca-Volmerange and Guiderdoni to reproduce the field number counts.
$\Phi^{\ast}_i$ is the normalisation of the LF for each morphological type
(Table \ref{tab-model}). In the integration over the luminosities, 
we stop the bright end of the blue LF at $M_{Bmax} = -23.5$ 
for any $z$ in model 1, and at $L_{Bmax}(z) = L_{Bmax}(z=0)/(1+z)^\eta$ 
with $L_{Bmax}(z=0)$ corresponding to $M_{Bmax} = -23.5$ for model 2.

\end{itemize}

Model 2 produces more faint
objects at high redshift than model 1, a trend which is
consistent with the results of the Autofib Redshift Survey (\cite{ellis96}).
Nevertheless, this qualitative model is somewhat unrealistic because the 
photometric properties induced by merging are not investigated in details.
In any case, model 2 is a useful approach to the $q_0=0.5$ scenario in
view of the high $M/L$ values found in weak lensing analysis 
(see \cite{narayan96} for a review) which seem to reject a low 
$\Omega$ universe. Finally, the number counts of arcs and arclets are 
limited to redshifts higher than the lens redshift, and they are mainly 
sensitive to rather bright objects, even at the faintest magnitudes.
For this reason, these counts do not depend severely on the uncertainties 
on the faint end slope of the local LF. 

\subsubsection{Surface brightness}
As the detection of extended objects is highly dependent on surface brightness,
we include this effect in the models. We assign a gaussian distribution of
central surface brightness in $B$ ($\mu^0$) to each morphological type. The 
parameters defining these observed distributions in the local
universe have been taken from the literature (all the units are in
$mag/\arcsec^2$):

\begin{itemize}
\item E: $<\mu^0> =17.55$, $\sigma=0.15$ (\cite{king78})
\item S0, Sa, Sb and Sc: $<\mu^0>$ $ = 21.02$, $\sigma=0.42$ (\cite{vdkruit87})
\item Sd and Im: $<\mu^0>$  $= 22.24$, $\sigma=0.49$ (\cite{vdkruit87})
\end{itemize}

The evolution of  $\mu^0$ with $z$ obeys to the standard relationship:
\begin{equation}
$$\mu^0(z)=\mu^0(z=0)+k(z)+e(z)+2.5 \log (1+z)^4$$
\end{equation}
where $k(z)$ and $e(z)$ are computed with the Bruzual \& Charlot code.
When no other value is indicated, the surface brightness limit for the 
detection is 28\, $mag/\arcsec^2$ in $B$, a value which is close to
the $2 \sigma$ detection level in most cases considered here. At redshift
$z$ and for each morphological type and magnitude, a visibility weight
is introduced according to the above surface brightness distributions
to compute the total fraction of objects actually detected. 

On ground-based images, the seeing spreads the light coming from the 
center of the objects. For number counts of arcs and arclets,
the effect of seeing is also important because it modifies the shapes
of the images (see details in \S 2.3 and \S 2.4). Several effects should be 
considered to properly take into account the
results of seeing on galaxy profiles and to include an
observational cut in surface brightness. In principle,
the luminosity profile for each morphological type
defines a characteristic radius which is related to the total 
luminosity of the galaxy, $L(z)$. This characteristic
angular radius $\theta$ is then convolved with the
seeing to give an {\em observed} central surface brightness 
$\mu^0_{obs}$. A cut in $\mu^0_{obs}$ induces a maximum redshift at
which the galaxy can be observed. This effect is minor for elliptical 
galaxies because $z_{cut} > z_{form}$ provided
that the surface brightness limit is faint enough, but it should
be much more sensitive for spirals. We have adopted here the crude 
weighting described above, as a function of $\mu^0_{obs}$, and 
the selection is often performed on WFPC HST images, where these
seeing effects are negligible in practice.
A more realistic treatment of source
profiles will require a fair knowledge of the morphological evolution
of galaxies, taking into account that morphology is expected to be
strongly wavelength dependent.  


\subsubsection{Results}
\begin{figure}
\psfig{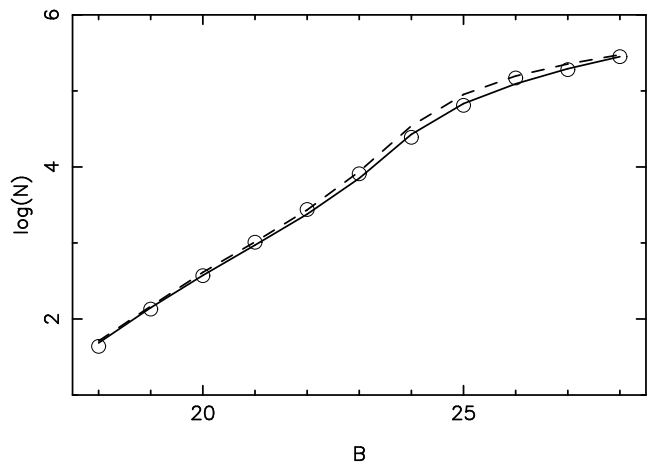}
\caption{Observed blue number counts ($\circ$) (from the compilation by
Metcalfe et al. (1991) completed by Williams et al. (1996)) and 
counts derived by the
model for model 1 (dashed line) and model 2 (solid line) per bin of one
magnitude and per square degree.
\label{fig-comptesB}
}
\end{figure}

\begin{figure}
\psfig{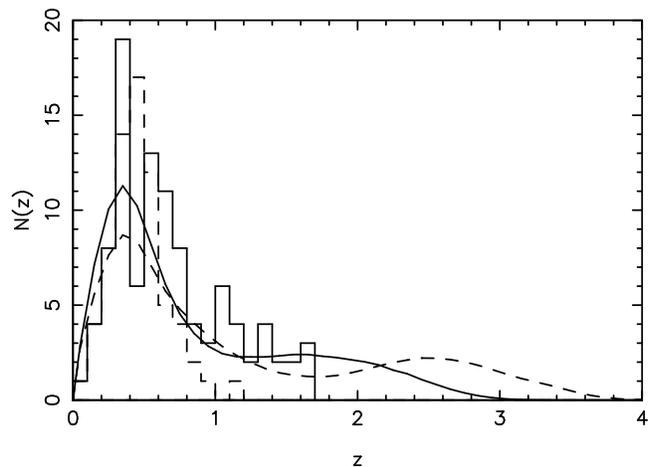}
\caption{Redshift distribution of galaxies with $22.5\leq B\leq 24$ for model 1 
(dashed line) and model 2 (solid line). Numbers are absolute counts per 
bin of 0.1 in redshift, normalized to the total counts by Cowie et al. (1996). 
The solid histogram is from Cowie et al. (1996) data and the dashed histogram is 
from Glazebrook et al. (1995). Models predict $\sim 20 \%$ of the total number
counts at $z \ge 1.5$, 
\label{fig-redshiftB}
}
\end{figure}

The predictions of field number counts in the $B$-band are presented 
in Figure \ref{fig-comptesB}, and the corresponding redshift distributions 
are shown in Figure \ref{fig-redshiftB}. All these calculations 
were performed without seeing.
Observed number counts are taken from the compilation by Metcalfe et al.
(1991) of various works. The counts at faint magnitudes were obtained 
by Williams et al. (1996) from the Hubble Deep Field survey. 
When we introduce a seeing of $0''.8$ and a more accurate 
treatement of galaxy profiles, only a slight difference appears for the 
faintest bin in magnitude, where the counts are reduced to $\sim 80 \%$
of their value without seeing.

We also checked that number counts in filters $U$ and $I$ were 
correctly reproduced as well as the redshift distribution of galaxies 
selected in $I$ ($17<I<22$) according to the CFRS (\cite{crampton95}).  
Infrared counts in the $K$-band are somewhat more discrepant and show an 
underestimate of the counts predicted at faint magnitudes. 


\subsection{Mass distributions in clusters lenses }
An accurate modelling of the mass distribution in the lens is needed because,
in principle, any variation in the local substructures or in the slope of the 
potential may induce a change in the local magnification and axis 
ratio of the background galaxies. Steep surface mass density profiles
will produce arclets narrower than a more gradual slope (\cite{hammer91}).
For this reason, the model is applied in the first place to two well studied 
cluster-lenses, using the published lens models: A370 at $z=0.375$ 
(\cite{kneib93}) and A2218 at $z=0.176$ (\cite{kneib95}, \cite{kneib96}).
Both clusters are very well represented by bimodal cluster-scale
mass distributions, with the clumps being centered on the two main galaxies
and the potentials modelled by pseudo isothermal elliptical distributions
(PIEMD, \cite{kassiola93}). In the improved modelling for A2218 (\cite{kneib96}),
the authors introduce an additional galaxy-scale component in the mass 
distribution. Each one of the 34 brightest cluster galaxies is modelled 
by a pseudo--isothermal elliptical mass distribution with the parameters 
(truncation radius, core radius and velocity dispersion) scaled to the galaxy 
luminosity. The effect on number counts of uncertainties in the mass distribution 
is quantified below (see \S 2.5). 
 
\subsection{Axis ratio and magnification}
The axis ratio of an arclet is the result of the intrinsic size of the source, 
tangential and radial magnifications and seeing. The
images from the Medium Deep Survey (the MDS project) show that the intrinsic 
size of galaxies seems rather constant until $z\simeq 0.8$ (\cite{mutz94}).
On the contrary, the results of 
lens inversion based on HST images (scanning higher redshift galaxies) 
lead to the conclusion that the half light radius of the sources of 
giant gravitational arcs is decreasing with increasing redshift following 
approximately the law $r_{hl} \propto 1/(1+z)$ (\cite{smail96}).
This effect is relevant in the merging hypothesis and implies
lower dimensions for high redshift objects.
We include the decreasing law as an additive hypothesis in models 1b
and 2b, for $q_0=0$ and $q_0=0.5$ respectively.
The evolution of the linear size then follows the relation
\begin{equation}
$$r_{hl}={8.7 \over 1+z} kpc$$ 
\end{equation}
scaled with the average half light radius of nearby spiral galaxies 
(8.7 kpc, \cite{mathewson92}), and also in good agreement with the
typical sizes observed for the most distant galaxies (\cite{trager97},
\cite{lowen97}). Models 1a and 2a are the equivalent ones 
with a constant linear size of 8.7 kpc, whatever the redshift.

All sources are considered circular because it is not worth introducing an 
ellipticity distribution when
we are only interested in arclets with
rather high axis ratio ($a/b\geq 2$), and when the intrinsic ellipticity of
sources is negligible compared to the one induced by the lens.
However, we are aware of the higher importance of source
ellipticities in weak lensing
studies, when reconstructing mass profiles. Hence the observed axis 
ratio is:
\begin{equation}
$${a\over b}=\sqrt{{(2 \lambda_t \theta_{hl})^2+s^2 \over 
(2 \lambda_r \theta_{hl})^2+s^2}}$$
\end{equation}
$\lambda_t$ and $\lambda_r$ being the tangential and radial magnification,
$\theta_{hl}$ the half light angular radius and $s$ 
the seeing or the width of the PSF.

\begin{figure}
\psfig{figure=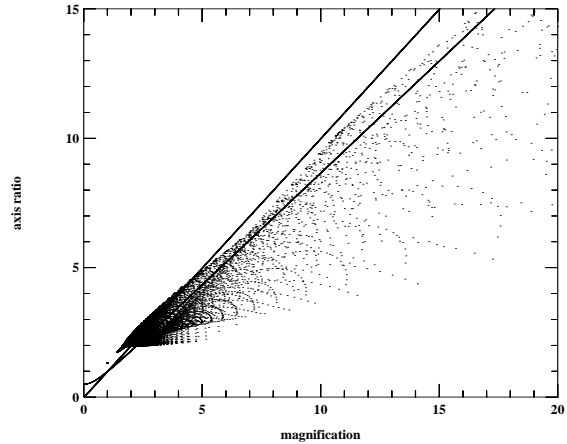,width=9cm}
\caption{Axis ratio versus magnification expected in the field 
of the HST image for background galaxies at $z$=0.8 
lensed by the cluster A2218, using 
the lens modelling by Kneib et al. (1995) and our model 1
(dots). For comparison, the same relation is displayed 
for a SIS potential with $\lambda_r=1$,
with and without a seeing of 1\arcsec (lower and upper lines
respectively). In ideal conditions, without either seeing or
pixel sampling, magnification and axis ratio are equal for a SIS.
\label{fig-axes2218}}
\end{figure}

$S(A,z)$ in equation 3 is the background surface at redshift $z$ where
the sources have an axis ratio between $A$ and $A+1$ after magnification. 
It is a specific term related to lensing and magnification,
which makes the arclet number counts quite different from results
in empty fields. Numerically, $S(A,z)$ is obtained by computing
the area in the image plane where arclets with 
$A\leq a/b \leq A+1$ can be found, and then dividing by the magnification
at each point because of the dilution factor in the image plane. 
The upper limit chosen for the magnification is 20, a
value representative of giant arcs, but of negligible effect
because statistics is not dominated by giant arcs.
In practice, as number counts will be computed in the image plane,
we have chosen to scan this plane with a grid of 1\arcsec\ step
and then to relate these image elements to the corresponding 
surface in the source plane at each redshift. 

In the case of a circular potential, two arclets at the same redshift with 
the same axis ratio would have identical magnifications. But 
the link between the axis ratio and the magnification of an arclet is not 
unique in the more realistic cases considered in this paper
(bimodal potentials, perturbing galaxies). 
Figure \ref{fig-axes2218} illustrates this effect on the relationship
between axis ratio and magnification, for a realistic set of parameters
in the cluster-lens A2218, compared to the results expected when the potential 
corresponds to a singular isothermal sphere (SIS). In practice, 
to compute the intrinsic magnitude of an arclet at redshift $z$ and
with an axis ratio between $A$ and $A+1$, 
the net magnification by the cluster was assumed to be the averaged 
magnification factor for all the arclets with the same axis ratio 
and redshift. 
The effect of the finite size of sources was neglected when computing
magnifications: 
in regions close to critical lines the magnification of an extended
source is different from that of a point source at the same location, 
it does not grow to infinity. Indeed, for magnifications smaller than 20 
this effect is quite negligible.
 
\subsection{Surface brightness effects}
In spite of the fact that gravitational lensing does not change the surface
brightness, other intervening effects such as atmospheric seeing, optical 
PSF and pixel sampling tend to modify it. Thanks to the magnification, an
arclet appears larger than the equivalent unlensed galaxy,
and its luminosity profile is stretched while the central surface 
brightness remains unchanged. Hence flattening by the seeing is less 
effective for a lensed object than for its source galaxy.
The relevant parameters become the characteristic lengths of the profile
along both axis of the arc, enlarged by the convolution with the full PSF.
In addition to simple magnification, gravitational lensing
makes the detection of faint galaxies easier by increasing their mean surface
brightness. This is even more sensitive for ground-based images (with seeing 
$\sim$ 1\arcsec ) as point sources are much more attenuated by the PSF. 
This effect can be approximated by considering the change in surface
introduced by the seeing:
\begin{equation}
$$\mu = \mu(\hbox{no seeing}) - 2.5 \, \log {a \, b \over \sqrt {(a^2 +s^2)
(b^2+s^2)}}$$
\end{equation}
where $a$ and $b$ are the semi major and semi minor axis of the arc and
$s$ is the seeing FWHM. In the case of HST data, the FWHM is equal to 
0.1 \arcsec\ although data are undersampled by the pixel size.   
Of course the result is also sensitive to the way $a$ and $b$ are measured,
the central part being more circular than the limiting isophote
after convolution. We choose here to consider the half light distances
in both axis. A more precise treatment is proposed by Hattori et al. (1997) who
consider isophotal magnitudes whereas we are concerned here with total
magnitudes, less sensitive to detection conditions. 
In the following we limit the counts to arcs with central surface brightness 
$\mu_B^0<26.5$ or $\mu_R^0<25.5$.

\subsection{Sensitivity to the mass distribution}
We check here the sensitivity of our model to
uncertainties on the mass distribution. In particular, we
focus on the effects of varying the assumptions for the
cluster potential in the two cluster-lenses considered:
A2218 and A370. 

\begin{figure}
\psfig{figure=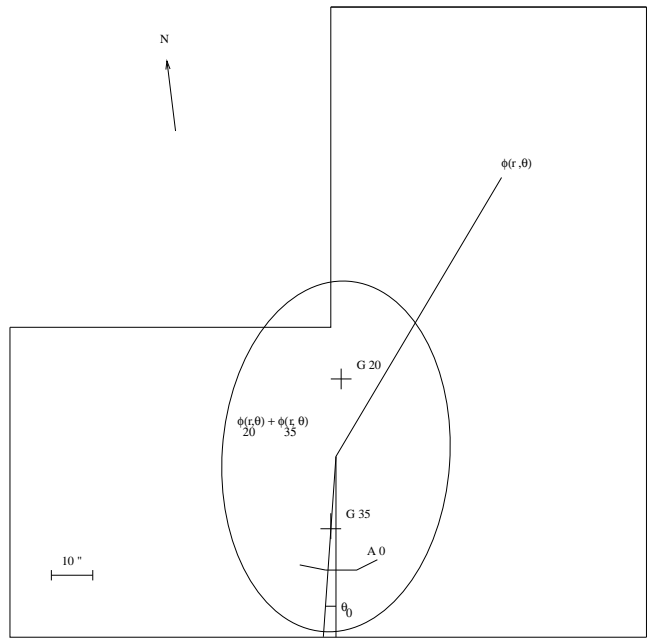,width=8.5cm,angle=-90}
\caption{Field diagram of A370 used in the simulations. The total size is
$ 2.5' \times 2.5'$ . The ellipse
separates the inner region, where the potential follows the expression of
Kneib et al. (1993), from the outer region where the slope is changed to check
the dependence of number counts on the shape of the mass distribution.}
\label{fig-potentielHST}
\end{figure}

\begin{figure}
\psfig{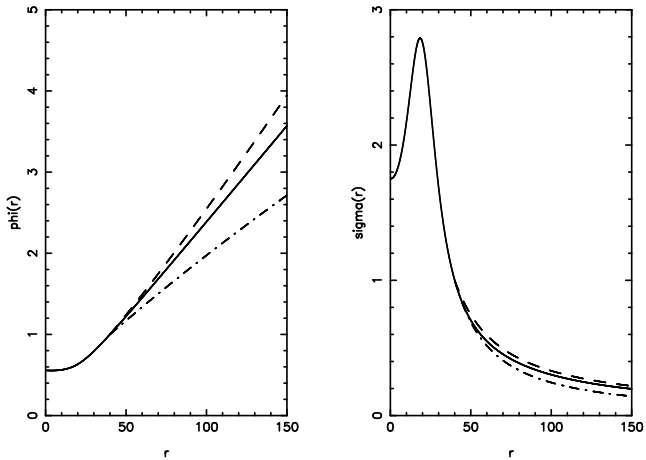}
\caption{Gravitational potential (left) and surface mass density (right) 
along the line G20 -- G35 in A370. Solid lines are for the expression given in 
Kneib et al. (1993); also shown are the cases: $\gamma=1.1$ (dashed line) 
and $\gamma=0.8$ (dot--dashed line).}
\label{fig-potentielprof}
\end{figure}

\begin{figure}
\psfig{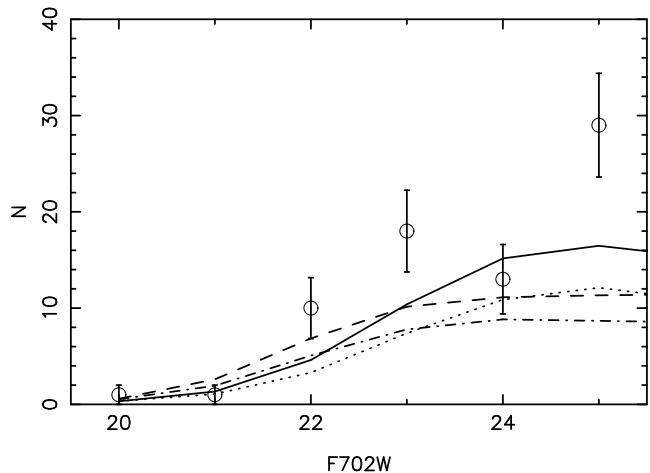}
\caption{Number counts of arclets in A2218 with the F702W filter, per bin 
of one magnitude, up to $R_{F702W}$=25. Selection criteria are: $a/b>2$ 
and $\mu^0_{F702W}\leq 25.5$. $\circ$: observed counts. The solid line 
corresponds to $q_0 = 0.5$ 
(model 2a) and the dashed line to $q_0=0.0$ (model 1a), both computed 
through the lens model by Kneib et al. (1996). The results obtained 
with the cluster-scale mass distribution only (Kneib et al. 1995) are also
displayed for comparison: dotted line and dot-dashed line correspond to
count models 2a and 1a respectively. No evolution in the 
source size has been included as it does not change the graphs (see text 
for more details). Errors bars correspond to statistical fluctuations.
}
\label{fig-histo2218}
\end{figure}

As explained in \S 2.2,
two lens models are available for the cluster A2218 (\cite{kneib95} and
\cite{kneib96}): the first one with only cluster-scale mass components,
and the improved one with 34 additional galaxy-scale components in the mass 
distribution. Both of them reproduce correctly the observed shear pattern.
Nevertheless, the second model increases the local magnification and distorts the
critical lines in the regions around each lensing galaxy, in such a way that
a lensed image located in this area will be divided in 2 or 3 parts whereas
the same object would appear as a giant arc with a smoother cluster-scale 
potential. Hence a higher number of arclets
is expected, for a given set of lens parameters (see also \cite{bartelmann95}). 
An example of such multiplication of arclets is given by
the giant arc \#359 at redshift $z=0.702$ (according to the numbering
scheme of \cite{leborgne92}). Introducing the local magnification by the mass
of galaxies \#341 and \#373 enabled to identify
three other images of this object (arcs \#328, \#337 and \#389). 
This point is quite crucial in our problem because although they do not 
dominate the arcs counts, multiple images can distort our statistics 
of small numbers, at least in a few magnitude bins. The difference 
between the expected number counts with these two models, compared to
real counts, is shown in Figure \ref{fig-histo2218}.

The cluster-scale model of A2218 (Kneib et al. 1995) has been used
to investigate the influence of uncertainties in the core radii and 
the velocity dispersions. To summarize, when these parameters are
modified within reasonable values (taking model uncertainties into account),
the net result is only a change in the absolute normalisation
of $N(z)$, with no shift in the mean redshift of the sample. 
The results are much more sensitive to errors in velocity dispersions 
than to core radii. When the core radii varies by 20\%, the number of 
arcs changes by only a 15 \%, whereas the total number of 
arclets is modified by 45\% for a small change of 10\% in the velocity 
dispersions. It is worth noting that such variations in the parameters are 
far beyond the published uncertainties.

The uncertainty introduced by the slope of the cluster potential in the 
external part of the deflector, where most of the arclets appear, has 
also been investigated with the lens model of A370. In this case, the
mass distribution in the central part of the cluster is very well constrained
by two systems of multiple images: the giant arc A0 and the pair B2/B3
lead to a bimodal mass distribution with two clumps centered on the
cluster galaxies G20 and G35 (\cite{kneib93}). Hence, we have introduced 
modifications in the external potential, keeping the center unchanged.
The external region 
is defined in Figure \ref{fig-potentielHST} by an ellipse oriented
parallel to the line linking the two central galaxies and with a semi major 
axis of 40\arcsec. Inside the ellipse we keep the potential to the
expression by Kneib et al. (1993)
\begin{equation}
$$\phi(r,\theta)=\phi_{G20}(r,\theta)+\phi _{G35}(r,\theta)$$
\end{equation}
where G20 and G35 are the main cluster galaxies (see \cite{mellier88}).
Outside the ellipse, the potential is not 
bimodal any longer but it follows the same law as for each clump:
\begin{equation}
$$\phi(r,\theta)=\phi_0 \left[\sqrt{1+\left({r \over r_c}\right)^2} +{\epsilon 
\over 2}
{({r \over r_c})^2 cos(2(\theta-\theta_0))\over \sqrt{1+({r \over r_c})^2}}
\right]^{\gamma}$$
\end{equation}
in polar coordinates,
with $\phi_0=6 \pi \left({\sigma \over c} \right)^2 r_c {D_{ls} \over D_s}$.
$\theta_0$ is the orientation, $\epsilon$ the ellipticity, $\sigma$
the velocity dispersion, $r_c$ the core radius,
$D_{ls}$ and $D_s$ are the angular diameter distances respectively
between the lens and the source and between the observer and the source. 
Note that this potential is continuous only along the major axis of the
ellipse in Figure \ref{fig-potentielHST}, but this simplification is 
only used to test the effects of the slope of the potential on the 
absolute normalisation of arcs number counts. 
Outside the ellipse, the parameters of the global potential are 
$\gamma=1$, $\epsilon=0.38$, $r_c=20 \arcsec$ and $\sigma=1000 $ $km s^{-1}$
to fit the modelling of Kneib et al. (1993).
The values $\gamma=0.8$ and $\gamma=1.1$ have been chosen to
provide respectively an underestimate and an overestimate of the surface
mass density of about 1/3 at large distance from the center of the
cluster (Figure \ref{fig-potentielprof}). Whatever the model used, 
the change in the amount of mass in the external parts of the cluster-lens 
simply scales the total number of arcs seen through a cluster, with a 
negligible change in the mean redshift of the sample ({\em i.e.} an 
overestimate of the total mass, $\gamma$=1.1, increases the number of arclets 
by 25 \%\, while an underestimate, $\gamma$=0.8, will decrease it 
by 30 \%). More details about the uncertainties induced by the mass 
modeling are discussed in the next section.

\begin{figure}
\psfig{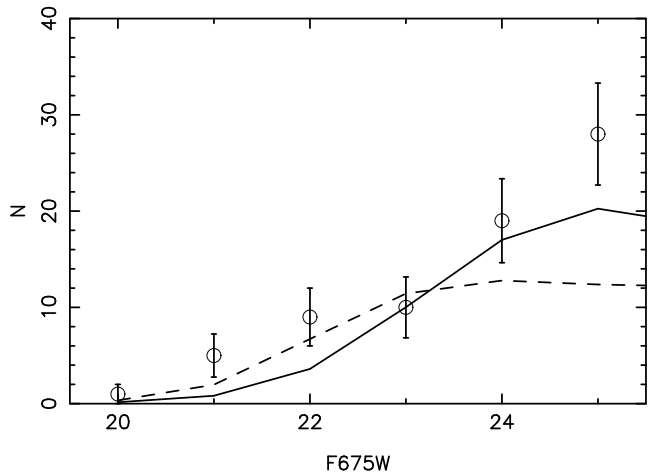}
\caption{Number counts of arclets in A370 with the F675W filter, per bin 
of one magnitude. Selection criteria are: $a/b >2$ and $\mu^0_{F675W}\leq 25.5$.
Same notations as in Figure \ref{fig-histo2218}.
}
\label{fig-histo370}
\end{figure}

\section{Results}
\subsection{Catalogs of observed arclets}
The detection of arclets in the two cluster-lenses 
was performed in the frame of WFPC2 HST images, obtained
from the STScI Archives: 5600 sec of exposure in filter F675W 
for A370 (P.I.: R.Saglia), and 6500 sec of exposure in filter 
F702W for A2218 (\cite{kneib96}). The $B-F702W$ color 
distribution of arclets in A2218 is also studied below. 
Blue magnitudes for A2218 were determined from ground-based deep B images
obtained a the 3.5m telescope  at Calar Alto (\cite{pello92}).
The pixel size was $0.25\arcsec$ and the seeing was $1.1\arcsec$
in this case.

The SExtractor package (\cite{bertin96}) was used to detect 
the arclets, with the requirement for an object to have at least 12 
contiguous pixels above $2 \sigma$ of the local sky level. 
The detection limit at 1$\sigma/pixel$ above
the sky is $R\simeq
25.2 mag/ \arcsec^2$ in A2218 and $R\simeq 24.9 mag/ \arcsec^2$ in A370.
We limit the sample to objects with total $R$ magnitude between
21.5 and 25.5 and axis ratio greater than 2, the bright
end cut is aimed to avoid contamination by cluster members.
A close inspection of each object has been done to eliminate
objects with problems in detection or photometry (close or inside the
PC, partially out of field,...), and also those showing position 
angles outside the range $\pm 45^o$ from the predicted local shear.
The final catalog contains 73 objects in A2218 and 81 objets in A370,
the number of objects excluded by weak shear constraints 
being 2 and 12 respectively.

\subsection{Absolute number counts}
The observed and predicted number counts of arclets versus R magnitude 
in A2218 are shown in Figure \ref{fig-histo2218}.
The predicted total number of arcs ($R_{F702W} \leq 23.5$ and $a/b\geq 2$) 
with the best mass model is lower than the observed number
by a factor of 2 (table \ref{tab-2218}). 
Similar counts in A370 with $R_{F675W} \leq 23.5$ and $a/b\geq 2$ are 
displayed in Figure \ref{fig-histo370}. Again, the models underpredict
the observed counts by a factor of $\sim 1.3 $ at the faintest magnitudes.
The contamination by cluster members is
possible, especially for the less elongated objects or the 
faintest ones. The number of objects excluded in each cluster
allows to estimate this source of contamination, which is unable to
explain all the observed excess. Some clues to understand this discrepancy
in terms of cosmological parameters, LF and clustering are given 
in \S 4.


\begin{table}
\caption[]{Comparison of different mass models for A2218 from
\cite{kneib95}
and \cite{kneib96}. Number of arclets with $R\leq 23.5$ and $\mu_R\leq
25.5$ in the frame of
the HST image for model 2a, compared to the observations.
}
\label{tab-2218}
\begin{flushleft}
\begin{tabular}{ccc}
\hline\noalign{\smallskip}
mass model&$a/b \geq 2$&$a/b\geq 3$\\
\noalign{\smallskip}
\hline\noalign{\smallskip}
bimodal (\cite{kneib95})&12.1&5.1\\
multimodal (\cite{kneib96})&16.7&7.9\\
observed arclets&30&13\\
\noalign{\smallskip}
\hline
\end{tabular}
\end{flushleft}
\end{table}

\subsection{Color distributions}

\begin{figure}
\psfig{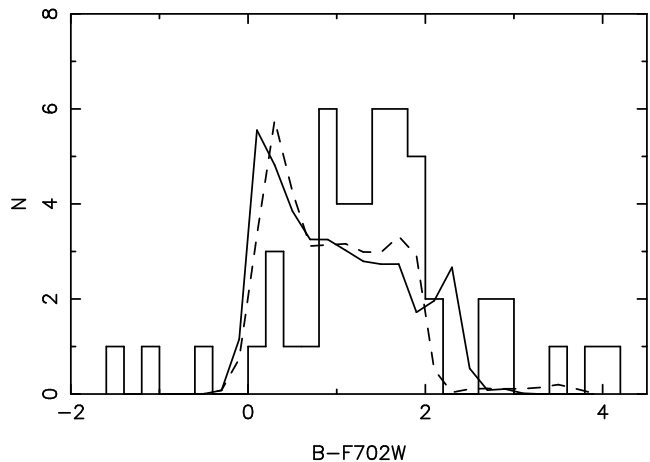}
\caption{Color distribution of arclets in A2218 with $R_{F702W}<25.5$ and
$a/b>2$ (mass modelling by Kneib et al. (1996)) for model 1a
(dashed line) 
and model 2a (solid line).
The histogram is the observed color distribution.
}
\label{fig-color2218}
\end{figure}

Another test of reliability for these results is to compare the
predicted with the expected color distribution of arclets.
To do that, the evolution with $z$ of the color index $B-F702W$ was
computed for the four morphological types of galaxy included in the model.
Then, these discrete values were replaced by a gaussian distribution
for each type ($\sigma=0.15$ mag.) and the counts in color were 
computed in the usual way. These predictions have been compared
to the observed values in A2218.
Because of the different
detection conditions in the $B$--band compared to the WFPC2 
(exposure time, seeing, pixel size), the resulting catalogue
is limited to 49 objects with $B<27.0$ and $a/b>2$,
21 of them at $B<24.5$.
The color distribution of arclets in A2218 for $R_{F702W}\leq 25.5$ 
and $a/b\geq 2$ is shown in Figure \ref{fig-color2218}.
The observed range of $B-F702W$ 
is well reproduced by the models, although the modelled colors
tend to be slightly bluer than observed. This 
effect is probably due to the way color indices are computed on CCD images: 
with such different sampling and observing conditions, colors are obtained
roughly as the difference between the measured magnitudes,
HST magnitudes being total ones while B-magnitudes are isophotal. 
As isophotal magnitudes tend to be overestimated at faint fluxes,
the net effect is an artificial reddening of the sample which could
explain this small discrepancy.

\subsection{Redshift distributions}

\begin{figure}
\psfig{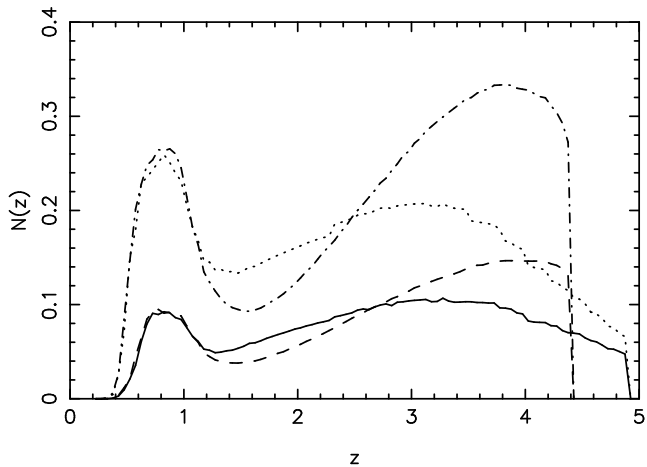}
\caption{Redshift distribution of arclets in A2218 per bin of 0.05 in $z$,
in the HST field ($R_{F702W}\leq 23.5$ and
$\mu_R \leq 24$) for $q_0 = 0$ (dot-dashed line:$a/b>2$, dashed
line: $a/b>3$) or $q_0 = 0.5$ (dotted line: $a/b >2$, solid line: $a/b
>3$).
}
\label{fig-z2218-R}
\end{figure}

\begin{figure}
\psfig{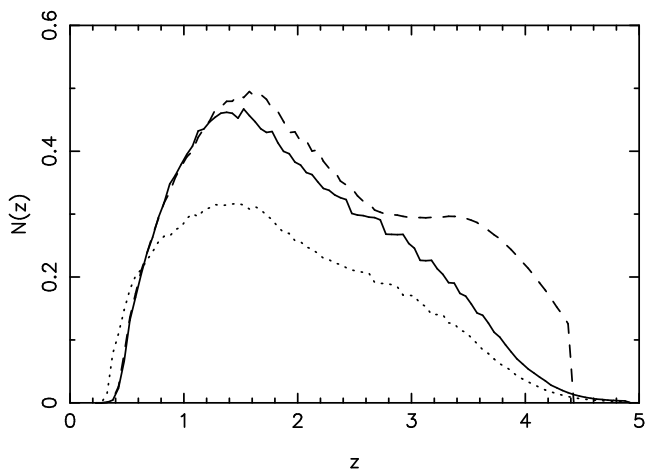}
\caption{Redshift distribution of arclets in A2218 per bin of 0.05 in $z$,
with $B\leq 24.5$ and $a/b\geq 2$, according to the mass distribution by 
Kneib et al. (1996) for
model 1a (dashed line) and model 2a (solid line).
Dotted line corresponds to
the bimodal mass distribution of Kneib et al. (1995) for model 2a.
The cut at $z=4.5$ corresponds to the redshift of formation for $q_0=0$.
}
\label{fig-z2218}
\end{figure}

\begin{figure}
\psfig{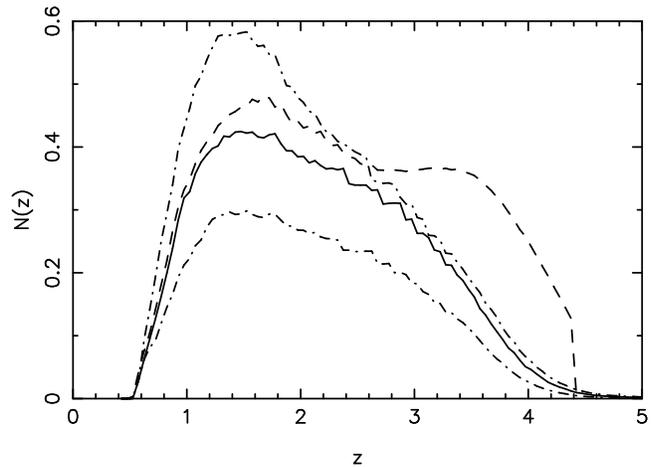}
\caption{Redshift distribution of arclets in A370 per bin of 0.05
in $z$ ($B\leq 24.5$ and $a/b\geq 2$) for 
model 1a (dashed line) and model 2a (solid line).
The dependence on the potential slope is shown as dot--dashed 
lines for $\gamma$=1.1 (upper line) and $\gamma$=0.8 (lower line) for
model 2a. 
The cut at $z=4.5$ corresponds to the redshift of formation for $q_0=0$.
}
\label{fig-z370}
\end{figure}

For a direct comparison with existing deep redshift 
surveys, the redshift distribution of arclets has been computed 
in both $B$ and $R$ filters. 
These results are shown in Figures \ref{fig-z2218-R} and \ref{fig-z2218} for
A2218, and in Figure \ref{fig-z370} for A370.
In the case of A2218, our results
can be partially compared to the results of a successful spectroscopic 
survey of arclets (\cite{ebbels97}), where 19 redshifts were 
obtained ranging from $z$ = 0.45 to 2.5, and observed magnitudes from 
R = 20.0 to R = 23.5. This observed distribution peaks to a mean value of 
$<z>=0.7$, with only two objects at $z>1$. Figure \ref{fig-z2218-R}
shows the redshift distribution of arclets expected with the same
selection conditions than that adopted by \cite{ebbels97}.
The peak observed is remarkably well reproduced. However, another
population of objects is expected at higher redshifts, $z>2$, 
with models 1 and 2, which is not seen in these data. This high
redshift tail is mainly produced by E type galaxies and it is
extremely sensitive to the redshift of formation assumed for the
sources and to the hypothesis of no-evolution in morphological types.
The maximum value of N(z) at $z>2$ is lowered by $30-50 \%$ 
assuming a redshift of formation $z_f \sim 6$, but the distribution
extends to higher redshifts keeping the total number of objects 
approximately constant.
Selection biases affecting the spectroscopic sample, the 
accuracy of the lens model and the effects of clustering behind the
lens are discussed below (\S 4).

We have also tried to compare our model with a larger sample of giant
arcs and arclets originating from different clusters (Figure
\ref{fig-zarcs}), from a compilation of all measured redshifts at the
present day. Again, in
this sample arclets are mostly found between $z$=0.5 and 1, although 
arcs produced by different mass distributions are mixed and cannot be
used for a close comparison with Figure \ref{fig-z2218}.
This sample is however interesting because it
corresponds to arcs with high magnifications so the integrated flux
favours the acquisition of better S/N ratio on the continuum of the
spectra. Some absorption lines are expected to be observed at least in a 
few cases of star forming galaxies, and they would help in principle 
to determine redshifts in the range $1.2 \le z \le 2.2$ where no emission 
lines are present in the visible spectrum. This may be the case for the giant 
arc in Cl0024+17 (\cite{mellier91}) where a blue continuum is detected
with no emission lines, but no firm identification has been proposed
up to now. Beeing aware that objects with $1.2 \le z \le 2.2$
are systematically missed in spectroscopic surveys because of the lack
of strong spectral features in their visible spectrum, 
we find again a similar trend: present-day 
arcs spectrocopy reveals very few objects at $z>2$, whatever the 
selection criteria are.
 
\subsection{Size evolution of sources}
\begin{figure}
\psfig{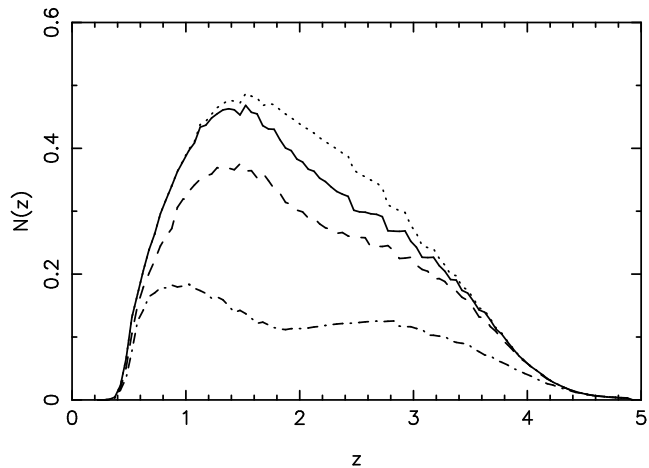}
\caption{Redshift distribution of arclets in A2218 per bin of 0.05
in $z$ ($B\leq 24.5$,
$a/b\geq 2$ and $\mu^0_B\leq 26.5$).
Solid line is for no seeing,
dashed line corresponds to a seeing of 1$\arcsec$ while dotted line
corresponds to no seeing and no threshold in surface brightness, all
curves are for model 2a.
Dot-dashed line corresponds to model 2b with a seeing of 1$\arcsec$ 
(evolution of the source size). 
}
\label{fig-z2218_seeing}
\end{figure}

\begin{figure}
\psfig{figure=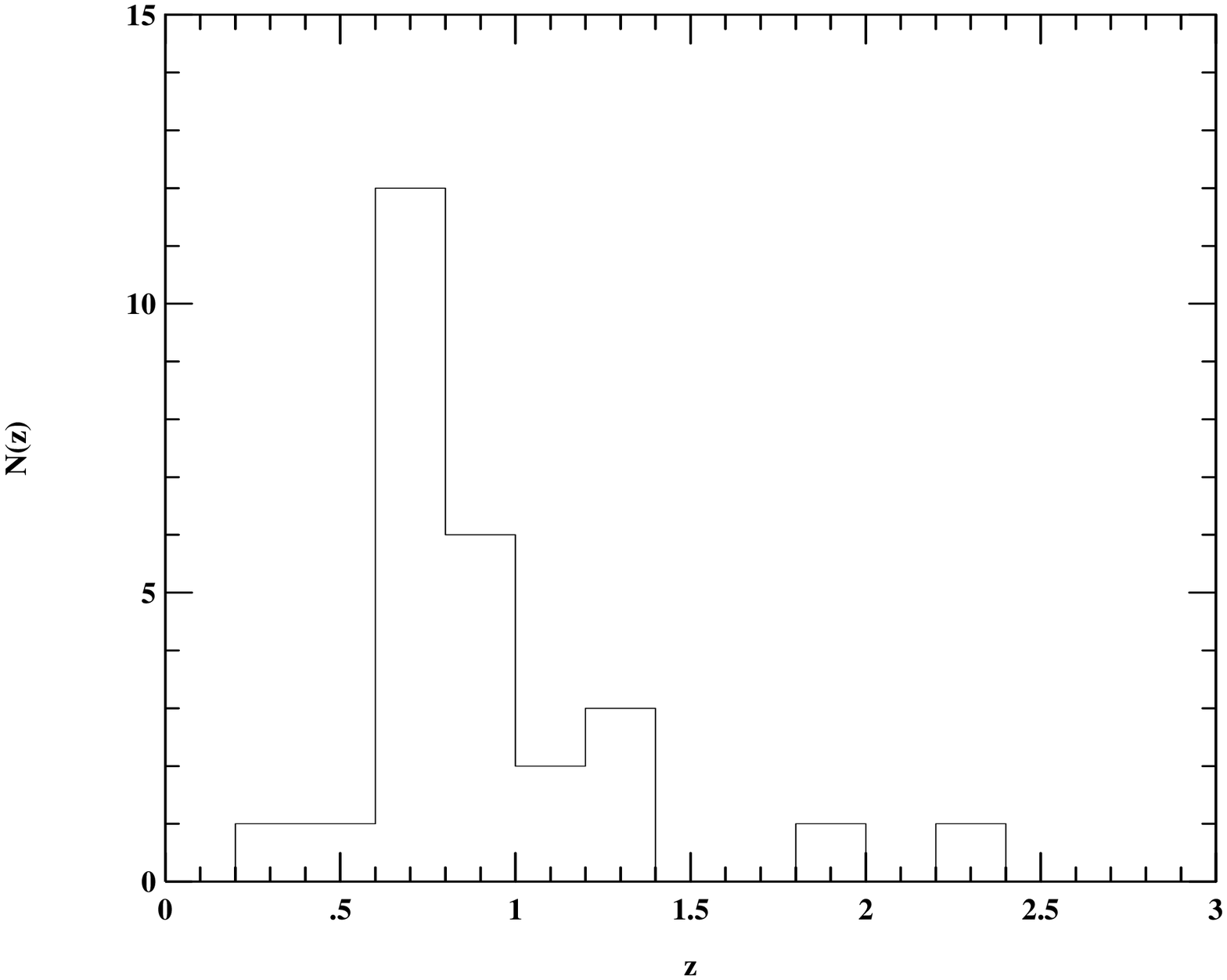,width=8cm,angle=-90}
\caption{Redshift distribution of arclets in several clusters except A2218.
References: 
\cite{soucail88} (A370), \cite{mellier91} (A370 and Cl2244--02), \cite{fort94} 
(A2163 and S295), \cite{pello91} (A2390), \cite{bezecourt97} (A2390), 
\cite{melnick93} (Cl2236--04), \cite{kneib94b} (Cl2236--04), 
\cite{smail95} (AC114), 
\cite{ellis91} (A963), \cite{lavery93} (GHO2154+0508), \cite{allen96} (PKS0745).
}
\label{fig-zarcs}
\end{figure}

When neither the seeing nor the surface brightness are 
considered, the total number counts and the color distributions 
of arclets are almost independent of the size evolution of sources 
with redshift. On the contrary, 
when realistic ground-based detection conditions are 
introduced in the model, the resulting redshift distribution 
can differ significantly.
Figure \ref{fig-z2218_seeing} shows these effects on the
redshift distribution of arclets in A2218. 
When an atmospheric seeing of 1$\arcsec$ is introduced
together with size evolution of galaxies,
the observed population splits into two
components: one corresponds to spirals at redshifts of the order of 1
and the other one to ellipticals at $z$ greater than 2.
The former probably constitutes the bulk of the redshifts
compiled in Figure \ref{fig-zarcs} because most of these arcs
were selected on ground based images and display emission lines. 
As expected, including an evolution of the size of galaxies
does not change the results on HST data where the PSF is always smaller
than the angular size of the sources whatever the redshift. 

\subsection{Optimisation of the search for high-z galaxies}
\begin{figure*}
\centerline{
\psfig{figure=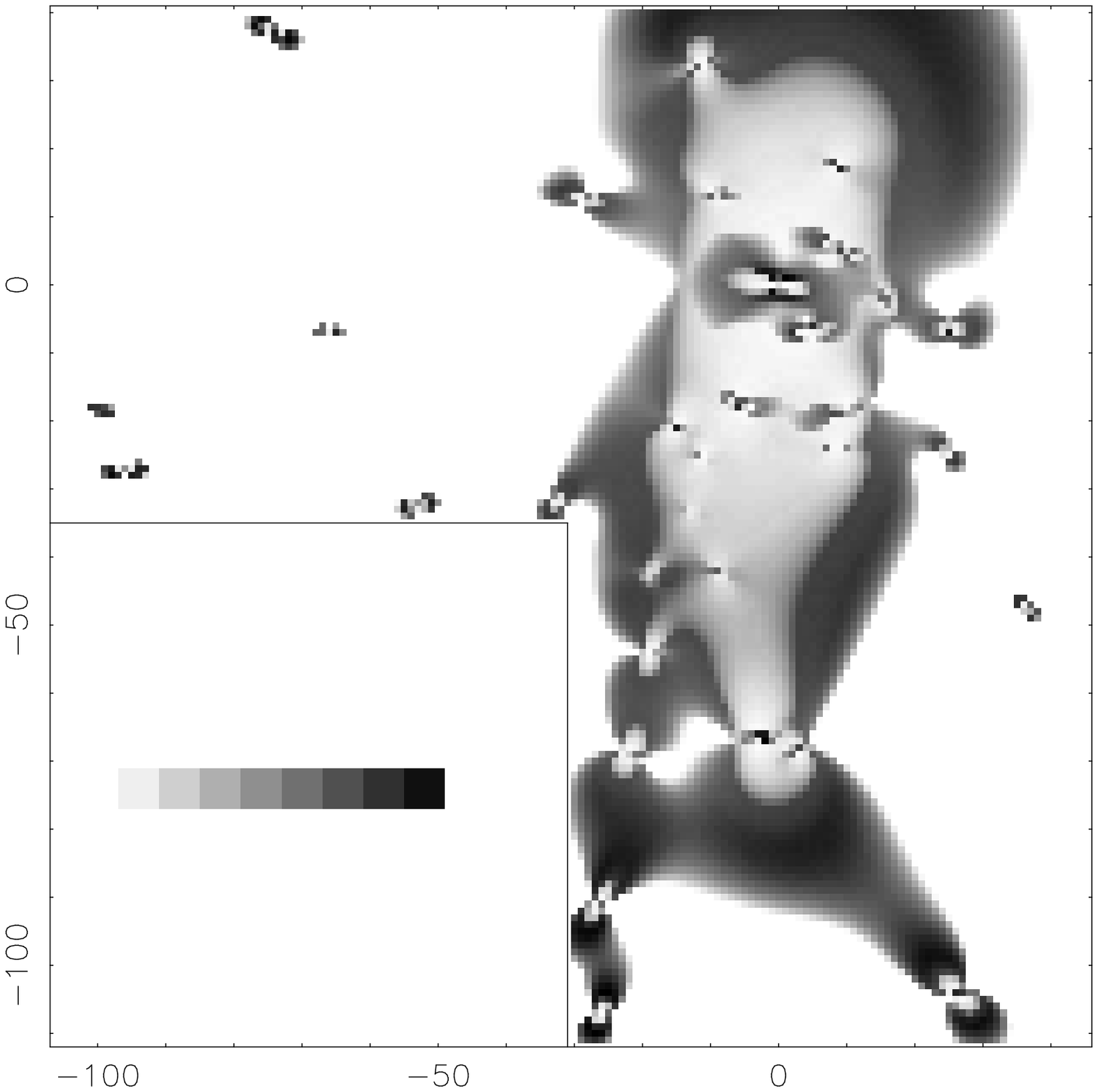,width=11cm,angle=-90}
\hskip -2cm
\psfig{figure=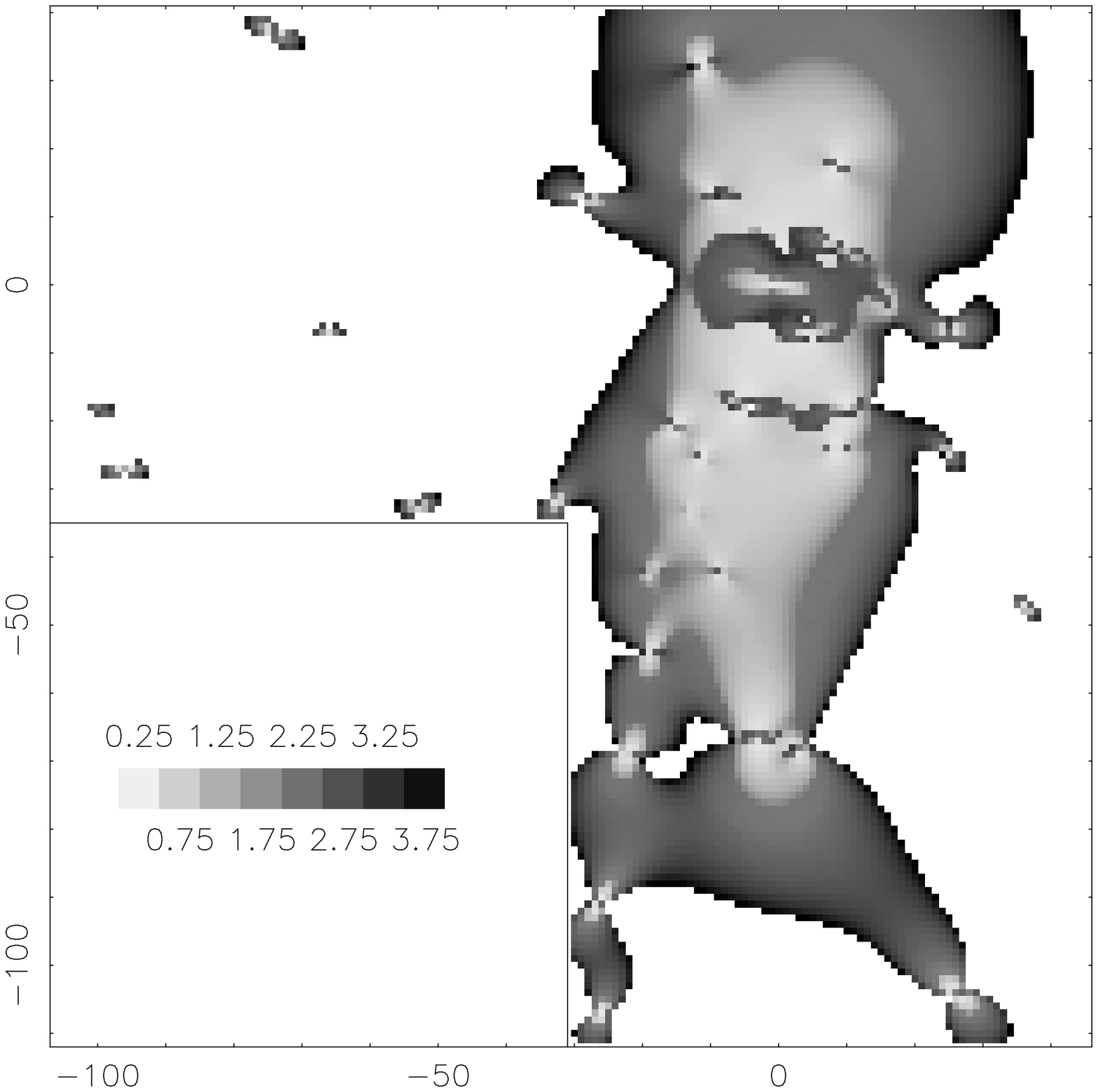,width=11cm,angle=-90}
}
\caption{Left: Density of arclets in cluster A2218 with $B<24.5$ and $a/b>3$
for model 1a. The arclets density increases from white
pixels (null density) to black pixels. Right:
Mean redshift of arclets in cluster A2218 with $B<24.5$ and $a/b>3$
for model 1a. Gray scale levels corresponds to redshift bins of $\Delta
z=0.5$ from [0.0,0.5] (clearer pixels) to [3.5,4.0] (black pixels).}
\label{fig-carte}
\end{figure*}

One of the most interesting issues of this work is to produce a
tool to select lensed galaxies at various redshifts.
Instead of computing counts in the whole field, one can compute 
the 2D distribution for both counts and mean redshifts, in order to estimate
and to compare the local densities all over the field.
For computational reasons, the surface brightness distribution
in \S 2.1.3  was replaced by the mean value of $\mu^0_B$ for each
morphological type. 
The results for A2218 are shown in Figure \ref{fig-carte}. 
The area where arclets are observed is fairly well identified (see
Figure 1 in \cite{kneib96} for comparison). 
These figures show the places where the numerous high redshift objects
(seen in the $N(z)$ curves) are expected. As one goes away from the cluster
center, only arcs with high $z$ can be lensed to acquire an axis ratio
greater than 3. However, in the limiting area where arcs can be found, the
cutoff in the density of arcs is quite sharp.
In the very center of the cluster, radial arcs at high
redshift are also predicted by the model.
These objects remain unobserved because of obscuration by the cD
enveloppe.

With the help of our model and maps similar to those in Figure \ref{fig-carte} 
we can optimize the search for well defined samples of arclets and/or 
the search for high redshift lensed galaxies. The selection procedure 
could even benefit from the combination with redshifts estimated through
multi-color photometric techniques (\cite{pello96}) or with the 
so-called ``lensing redshift'' (\cite{kneib94a}, \cite{kneib96}). Note 
that some high-z lensed galaxies at $z \ge 4$ have already been detected, although 
serendipitously, in the cluster Cl0949+4713 (\cite{trager97}), or in 
Cl1358+62 (\cite{franx97}), all of them showing "red" colors in the 
visible. This seems to indicate that, in the absence of a large wavelength
coverage, the selection will have to rely on morphological information
first and not on the usual criterium based on blue colors.

\section{Discussion}

The observed absolute number counts of arclets presented in
Figures \ref{fig-histo2218} and \ref{fig-histo370} tend to be
overestimated with respect to the predicted values, especially
at faint magnitudes where the excess attains a factor of 
1.3 to 2, depending on the cluster. It is worth noting that
computing number counts is difficult because of the 
large number of uncertainties involved in models, and some of them 
concerning the evolution of galaxies are quite similar to those 
encountered in empty fields. Firstly, 
counts in cluster lenses depend on the local normalisation
of the LF, which has an uncertainty of a factor of 2 
(see the discussion in \cite{ellis96}), 
partly due to the statistical fluctuations from field to field and 
to the clustering of background galaxies. 
Secondly, as we are looking 
deeply in only a few lines of sight, these fluctuations may introduce 
a bias because the 
cluster-lenses selected at first are among those with the highest 
number of arcs and arclets. This particular bias is difficult to
avoid, but it is not expected to induce a difference larger than
a factor of $2$ in the case of A2218, where the differences 
between observed and predicted number counts are the highest.
A possible excess of galaxies at $z \simeq 0.45$ is already mentionned
by Ebbels et al. (1997) but it does not seem to strongly distort the 
final redshift distribution. 
An example of clustering of background galaxies is observed
behind another cluster-lens, namely the system of arclets in A2390 where two
redshift planes are suspected behind the cluster (\cite{bezecourt97})
preventing to do this kind of analysis. 

Anyhow, a spectacular change in the number counts is obtained when we take 
into account the effects due to galaxy-scale mass components. 
The number of lensed objects is increased thanks to
both the local increased magnification and the additional
critical lines which divide giant arcs in smaller ones. A piece of 
evidence for this effect is given in \S 3.2 and \S 3.4 
in the case of A2218, where the number of arclets is a factor of 1.5 higher 
when galaxy-scale mass components are introduced (Kneib et al. 1996).
This is a strong justification of the need for accurate modelling when 
computing arcs statistics, and in this sense the present model is a clear 
improvement with respect to
other previous similar works (\cite{nemiroff89}, \cite{grossman94}). 
In particular, the absolute normalisation of counts through a cluster-lens
is approached in a reliable way for A2218. In the case of A370,
we plan to improve the lens modelling by adding the local effects of
individual galaxies as additional mass components, thanks to the
constraints given by new spectroscopic results 
which clearly confirm some multiple image candidates (B\'ezecourt et al. in
preparation). If
the results are similar to those found in A2218 as expected, this procedure 
may reconcile the observed number of arclets presently in excess
with model predictions for this cluster.

Another major point to discuss is the sensitivity of our results to the
geometry of the universe. The two models considered here have been
fixed to reproduce both number counts and redshift distributions of
galaxies in the field. The mass models used in all the
simulations were determined for a universe with $\Lambda =0$ and 
$\Omega =1$. It was assumed here that the same models were
still valid for $\Omega=0$ because varying $\Omega$ would not 
change the mass distributions by more than a few percents. 
This remark can also be extended to the value of $\Lambda$,
which has a negligible effect on the lens-model parameters
(less than $10 \%$), but a more significant impact on the 
expected distribution of background sources. 
As shown in Figures \ref{fig-z2218-R}
and \ref{fig-z2218}, the value chosen for $q_0$ affects the expected 
distribution of background sources at high redshift. 
The total number counts are also sensitive to $q_0$ 
(Figures \ref{fig-z2218-R} and \ref{fig-z2218}),
especially at faint magnitudes where the main contribution comes
from the more distant objects. 
Roughly speaking, when $q_0$ changes from $0.5$ to $0$ 
keeping $\Lambda = 0$,
the efficiency of the cluster-lens given by N(z) increases by 
a small factor of less than $\sim 10 \%$ up to $z \sim 3$,
but it rises dramatically up to a factor of $\sim 200 \%$ at the
highest $z$ (Figures \ref{fig-histo2218} and \ref{fig-histo370}).
The maximum effect on absolute number counts is $\sim 30 \%$
for the faintest bins in magnitude ($B \sim 27-28$). 
When we change from a matter dominated ($(q_0, \lambda_0) = 
(0.5, 0.0)$) to a $\Lambda$ dominated geometry
($(q_0, \lambda_0) = (0.0, 0.5)$) keeping $\Omega =1$
($\Omega = 2q_0 + 2\lambda_0$),
the predicted number counts of arcs will increase by a 
$15 \%$ even at moderate magnitudes ($B=24-25$). 
According to this result, models with $\lambda_0 \geq 0$ could
reconcile observed with predicted number counts of arcs.
The increase in number counts is even higher when we consider
models with low values of $\Omega $: it attains $40 \%$ 
at the same moderate magnitudes with $\lambda_0 \sim 0.9$ and 
$q_0 \leq 0$, but this model is somewhat unrealistic.

The main discrepancy between predicted and observed N(z) for arclets is
the apparent lack of objects observed at $z \ge 1$. Several arguments 
can be proposed to account for these missing galaxies:
(a) a systematic bias in the sample of arclets selected for spectroscopy
and/or in the successful sub-sample, 
(b) lens modelling is not accurate enough, and (c) the
spectro-morphological properties of 
galaxies above $z\simeq 1$ are not well represented by
the evolutionary models adopted here.

(a) Objects with $z$ between about 1.2 and 2.2 can be missed in
spectroscopic surveys because of the lack of any emission lines in the 
visible part of the spectrum. 
In the spectroscopic sample of Ebbels et al. (1997), it is striking that all the
redshift determinations considered as secure correspond to spectra with
an emission line, generally identified with [OII]. This clearly
indicates a bias in the spectroscopic sample to redshifts lower than 1.2
or higher than 2.2. About $40 \%$ of this sample of arclets 
remains with no redshift determination and may correspond to
galaxies at $z>1$. A more detailed examination of their 
photometric SEDs might allow to deduce a photometric redshift
and to discuss this particular point. 
Anyway, as far as no well defined and
magnitude-limited samples of arclets are studied yet, no direct
comparison with our work can be proposed. 
 
(b) Sensitivity to the lens mass distribution has been investigated by 
several methods, and we have shown that only minor effects apply on the 
redshift distribution when we 
change the main parameters of the lens models such as the core radius, 
the velocity dispersion or the slope of the potential, although this 
can significantly change the total number of lensed objects. 

(c) Several uncertainties remain in the general problem of modelling 
galaxy evolution which could modify the predicted N(z). 
Computing the redshift distribution of gravitational arclets with $B\leq 
24.5$ is in fact equivalent to the distribution of field galaxies up to
$B\simeq 26-26.5$ or fainter after magnification by the cluster. As
the evolution model is constrained roughly until to $B=24$, where 
spectroscopic data for field galaxies are available (\cite{glazebrook95}, 
\cite{cowie96}), a discrepancy may appear at fainter levels. 
Hence, we would meet here once
more the excess in the number of high--$z$ objects initially found in
works about numbers counts (\cite{bruzual80}, \cite{tinsley80}, 
\cite{guiderdoni90}, \cite{metcalfe91}). 
It should be noted that the introduction of internal absorption by dust 
in the model
could decrease the counts at high redshift as we are mostly
interested by the $B$--band, equivalent to the rest frame $UV$-band,
where absorption effects are more important. This hypothesis has been 
explored by Campos and Shanks (1995). Metallicity effects
are not taken into account here (metallicity is assumed to be solar),
and would act in the opposite way because the UV luminosity is
increased for a lower metallicity (as expected at earlier epochs).

The last point to comment is the difference found in the N(z)
distributions behind the two clusters. The fraction of galaxies 
at $z<1$ is quite small in the case of A370 with respect to the high 
redshift tail. This is essentially related to the difference in the 
redshift of the lens: the magnification of galaxies at $z<0.6$ is 
much less efficient in A370 ($z_{lens}=0.37$) than in A2218 
($z_{lens}=0.17$). This is also visible in the spectroscopic 
redshift survey of A2218 where 50\% of the redshifts are smaller 
than 0.6. 

\section{Conclusions}
We have shown that detailed absolute number counts, color and
redshift distributions can be computed for lensed galaxies 
through an accurate modelling of the cluster-lens mass
distribution. The framework for galaxy evolution has been chosen
to fairly reproduce the observed number counts and redshift 
distribution of field galaxies. The interest in applying these
calculations to arclets is to use cluster-lenses as 
filters to select faint distant galaxies. We have applied this
procedure to two different cluster-lenses, A2218 and
A370, for which the mass distribution is fairly well known,
and we have studied the impact of the different sources of
uncertainty on the predicted number counts and redshift 
distributions, taking into account the observational conditions. 
The main result is that arcs at redshifts between 0.5 and 1 are 
correctly predicted by the modelling as observed. Nevertheless, 
an important population of high redshift arclets ($z \ge 1.0 $) is also 
revealed by the simulations, which is not observed in spectroscopic
surveys of arclets. This disagreement could
result partly from a bias in the spectroscopy of arclets, but the 
main contribution is probably due to
uncertainties in the evolutionary models for galaxies at high
redshift. 

In summary, our results show that a detailed model for the
cluster-lens, including galaxy-scale mass components, is absolutely
needed to interpret the observed distribution of arcs and arclets
in terms of general properties of the background population of
galaxies. A good agreement between model and observed absolute 
number counts can be obtained by a fine tuning parametrization
of the evolutionary models for galaxies and/or the cosmological
parameters. In this respect, the present work joins the problems
encountered in the modelling of faint field galaxy samples
(PBZ, \cite{rocca90}). Nevertheless, the difference in this case
is that we are selecting high redshift galaxies and rejecting the
faint neighbouring population. As evolutionary effects are
extremely sensitive to the behaviour of sources at high redshift,
the more distant the lens is, easier it will be to constrain them,
provided one is able to detect a cluster lens with a sufficient
number of arclets. In a more prospective way,
observing distant cluster-lenses
could help on disentangle the role of pure geometrical effects,
giving constraints on the world model, from pure spectromorphological
evolution. In any case, the piece of work presented here is just 
the beginning of a study, an example on two single clusters at moderate
redshift, but the effort has to be pursued on a complete sample 
of lenses in order to minimize the fluctuations and possible
clustering along the line of sight, which is difficult to avoid.
For this reason, it is essential to define a 
homogeneous sample of cluster-lenses in order to derive 
reliable constraints on the absolute number of background galaxies.
This could be explored by using the sample of distant 
clusters selected from the EMSS catalog (\cite{gioia94}), but it would require 
an analysis of HST images for each cluster combined with an accurate modelling 
for all those identified as cluster-lenses. 

Another interesting effect which is amplified when selecting high
redshift objects is the role of elliptical galaxies. They are
responsible for the bulk of the presently undetected population of
high redshift arclets at $z \ge 2$. Changing the redshift of 
formation modifies the expected N(z) distribution at such redshifts, 
but it does not reduce significantly the total excess in number. 
The most straightforward way to solve the problem is to break down 
the hypothesis that ellipticals form in a unique burst. This idea
is also supported by the deep HST images (see also a discussion in
\cite{baugh97}), where the distant
galaxies seem rather irregular, deviating from the pure elliptical shape
assumed here. In a more general way, there is no observational reason
to assert that the progenitors of present day galaxies follow the simple
evolutionary law used here (spectrophotometric evolution + $(1+z)^4$ dimming
in surface brightness), and the first results on the arclet sample
strongly support this idea. There is more probably a strong relationship 
between morphology and spectrophotometry, the two aspects being both 
interdependent and wavelength dependent. In any case,
the identification of
the distant progenitors of ellipticals remains an exciting challenge.

Computing the 2D distributions in number counts or mean redshift
seems to be an important tool to build up independent samples of 
high redshift galaxies. We propose to couple together the computational
tool presented here with photometric redshift techniques in order
to select the spectroscopic samples. It is worth noting that arcs
and arclets correspond to a sample of galaxies much less affected 
by biases in intrisic luminosity. Besides, the brightest 
galaxies at any (high) redshift will be seen through lens
magnification before being detected somewhere else.

\acknowledgements

We thank P'tit Lu Van Waerbeke, JP. Kneib, Y. Mellier, G. Bruzual and 
B. Fort for useful discussions and helpful comments. This work was partly
supported by the Groupe de Recherche Cosmologie and by the French
Centre National de la Recherche Scientifique.

\end{document}